\documentclass[english,aps,manuscript,superscriptaddress]{revtex4-1}
\usepackage[T1]{fontenc}
\usepackage[latin9]{inputenc}
\usepackage{verbatim}
\usepackage{amsmath}
\usepackage{amssymb}
\usepackage{graphicx}

\usepackage{color}

\makeatletter

\providecommand{\tabularnewline}{\\}

\makeatother

\usepackage{babel}
\begin{document}

\title{In-plane transverse polarization in heavy-ion collisions}

\author{Anum Arslan}
\email{anumramay@mail.ustc.edu.cn}
\affiliation{Department of Modern Physics and Anhui Center for Fundamental Sciences
in Theoretical Physics, University of Science and Technology of China,
Hefei, Anhui 230026, China}

\author{Wen-Bo Dong}
\email{wenba@mail.ustc.edu.cn}
\affiliation{Department of Modern Physics and Anhui Center for Fundamental Sciences
in Theoretical Physics, University of Science and Technology of China,
Hefei, Anhui 230026, China}
\affiliation{Institute for Theoretical Physics, Goethe University, Max-von-Laue-Str. 1, D-60438 Frankfurt am Main, Germany}

\author{Charles Gale}
\email{gale@physics.mcgill.ca}
\affiliation{Department of Physics, McGill University, Montreal, Quebec H3A 2T8,
Canada}

\author{Sangyong Jeon}
\email{jeon@physics.mcgill.ca}
\affiliation{Department of Physics, McGill University, Montreal, Quebec H3A 2T8,
Canada}

\author{Qun Wang}
\email{qunwang@ustc.edu.cn}
\affiliation{Department of Modern Physics and Anhui Center for Fundamental Sciences
in Theoretical Physics, University of Science and Technology of China,
Hefei, Anhui 230026, China}
\affiliation{Department of Physics, McGill University, Montreal, Quebec H3A 2T8,
Canada}

\author{Xiang-Yu Wu}
\email{xiangyu.wu2@mail.mcgill.ca}
\affiliation{Department of Physics, McGill University, Montreal, Quebec H3A 2T8,
Canada}

\begin{abstract}
We give an analytical expression for the in-plane polarization $P^{x}$,
in heavy-ion collisions that has, to our knowledge,
not been measured in heavy-ion collision experiments.
We also carry out a numerical study of $P^{x}$ using a hydrodynamic
model simulation as a cross-check for the analytical formula.
It is found that if the temperature-gradient contribution is neglected the simulation result
for $P^{x}$ qualitatively agrees with the analytical one.
The prediction of $P^{x}$ can be tested in experiments and will contribute to provide a complete and consistent
picture of spin phenomena in heavy-ion collisions.
\end{abstract}
\maketitle

\section{Introduction}

A substantial part of the orbital angular momentum in heavy-ion collisions
can be transferred into the strong interaction matter and leads to
the spin polarization of final-state particles through spin-orbit
coupling \citep{Liang:2004ph}.
This particular type of polarization is parallel to the normal vector of
the reaction plane formed by the impact parameter and the beam
direction which is fixed for all particles in one single event.
As such, it is usually referred to as the global spin polarization to distinguish it from
the spin polarization effects in proton-proton \citep{Bunce:1976yb,Heller:1978ty,Thomas:1978ova,Wilkinson:1981jy,DeGrand:1980gc,DeGrand:1981pe,Heller:1996pg}
and electron-positron collisions \citep{Belle:2018ttu}. There, the
spin polarization of particles is normally with respect to the production
plane spanned by the particle's momentum and the beam direction which
vary for particles with different momenta even in the same event.
Early studies of the global spin polarization can be found in
Refs. \citep{Liang:2004ph,Voloshin:2004ha,Gao:2007bc,Betz:2007kg,Becattini:2007sr}.

The global spin polarization of $\Lambda$ hyperons has been measured
across a wide range of collision energies \citep{STAR:2007ccu,STAR:2017ckg,STAR:2018gyt,HADES:2022enx,ALICE:2019onw}.
Driven by experimental progress, the global spin polarization has
been extensively studied in theoretical models.
In Ref.~\citep{Liang:2004ph}, a theoretical model was proposed in which unpolarized
quarks are scattered at fixed impact parameters by a static potential,
leading to polarized quarks after scatterings as the result of the
spin-orbit coupling. The model was later applied to multiple scatterings by the static potential \citep{Huang:2011ru}.
As a major improvement of the model,
a formalism for two-to-two quark scatterings at fixed impact parameters was developed \citep{Gao:2007bc}.
These results \citep{Liang:2004ph,Gao:2007bc} are only for one single
scattering. In a thermal system, particle collisions take place with
arbitrary incident momenta. Therefore, one has to take ensemble average over
all possible collisions in order to obtain an average effect \citep{Zhang:2019xya}.
In this way, it was shown that the spin polarization arising from
the spin-orbit coupling in one single scattering can be converted
to that from the spin-vorticity coupling after ensemble average over
all possible collisions in a thermal system \citep{Zhang:2019xya}.
In practice, spin polarization in hydrodynamical and transport models \citep{Xia:2018tes,Karpenko:2016jyx,Sun:2017xhx,Li:2017slc,Wei:2018zfb,Vitiuk:2019rfv,Becattini:2020ngo,Fu:2020oxj,Ryu:2021lnx,Jiang:2023vxp,Palermo:2024tza}
is calculated by mapping the vorticity to the spin polarization on the freeze-out hypersurface
\citep{Becattini:2013fla,Fang:2016vpj}.
For recent reviews on global spin polarizations, see Refs.~\citep{Wang:2017jpl,Florkowski:2018fap,Huang:2020dtn,Gao:2020lxh,Gao:2020vbh,Liu:2020ymh,Becattini:2022zvf,Hidaka:2022dmn,Becattini:2024uha}.

In addition to the global spin polarization, the spin polarization along the beam direction was also proposed in
hydrodynamic and transport models \citep{Becattini:2017gcx,Xia:2018tes} with the expected
behavior $P^{z}\sim-\sin(2\phi_{p})$ where $\phi_{p}$ represents
the transverse azimuthal angle of the hyperon's momentum
in the reaction plane. However the experimental measurement \citep{STAR:2019erd} shows
an opposite sign behaviour $P^{z}\sim\sin(2\phi_{p})$. The first theoretical explanation of this discrepancy
was provided in Ref. \citep{Wu:2019eyi,Wu:2020yiz}, where it was
found that the temperature vorticity qualitatively accounts for spin
polarization along the beam direction as well as the global spin polarization
as a function of $\phi_{p}$. Later on, it was found that the contribution
from the shear stress tensor can also yield the correct sign for longitudinal
polarization \citep{Fu:2021pok,Becattini:2021iol,Yi:2021ryh,Florkowski:2021xvy,Wagner:2022gza,Wu:2022mkr}.
More recently, it has been proposed that in a thermal model the projected
thermal vorticity along with dissipative corrections can describe
the behavior of longitudinal polarization \citep{Banerjee:2024xnd}.
Very recently dissipative relativistic spin hydrodynamics was developed from quantum kinetic theory for massive particles with non-local collisions \citep{Weickgenannt:2020aaf,Weickgenannt:2021cuo,Weickgenannt:2022zxs} and gave a good description of longitudinal polarization data \citep{Sapna:2025yss}.
These theoretical approaches collectively indicate that global equilibrium
has not been achieved, necessitating the inclusion of off-equilibrium
effects in the analysis \citep{Sheng:2021kfc}.

Surprisingly, there is a simple way to explain the behavior of the longitudinal polarization. If one uses the non-relativistic approximation in the blast-wave model \citep{STAR:2019erd,Voloshin:2017kqp}, one can show that
$P^{z}\sim\omega^{xy}\sim(1/r)v_{2}v_{r}\sin(2\phi)$
where the profile of the transverse radial flow velocity is given by $\mathbf{v}\sim\mathbf{e}_{r}v_{r}\left[1+v_{2}\cos(2\phi)\right]$, $\omega^{xy}\sim \nabla_{x}v^{y}-\nabla_{y}v^{x}$ is the longitudinal component of the vorticity vector, and $v_{r}$ and $\mathbf{e}_{r}=(\cos\phi,\sin\phi)$ denote the
radial flow velocity and its direction, respectively.
This provides a straightforward explanation for the experimentally observed pattern of the longitudinal
spin polarization. However, it is not clear whether such a simple
non-relativistic approximation \citep{STAR:2019erd,Voloshin:2017kqp}
could describe other spin observables without relativistic effects
that could potentially alter the observed pattern in experiments.

Inspired by the simple and intuitive blast wave picture of the longitudinal spin polarization,
we had performed a comprehensive analysis of spin observables in the framework of the modified or extended blast
wave model \citep{Retiere:2003kf,Jaiswal:2015saa,Yang:2016rnw,Yang:2020oig,Yang:2022yxa}.
We found analytical expressions or solutions for the longitudinal
and global spin polarizations as functions of particle's momentum
and collision centrality under flow-momentum correspondence \citep{Arslan:2024dwi}.
From the analytic solutions, one could clearly see that the global
spin polarization is driven by the directed flow, while the longitudinal
spin polarization is driven by the ellipticity in the emission zone and flow.
The analytical solutions can be improved systematically
by perturbative expansion in
the small deviations from the flow-momentum correspondence.

In this paper, we provide an analytical expression for the in-plane spin polarization $P^{x}$,
that has, to our knowledge, not been measured in experiments.
There are some earlier theoretical results for $P^{x}$ from hydrodynamical or transport model simulations
\citep{Karpenko:2016jyx,Xia:2018tes,Sun:2021nsg},
but the magnitudes of $P^{x}$ are very small compared with those of $P^{y}$.
Our analytical formulas show that $P^{x}$ and $P^{y}$ have the same order of magnitudes.
We also carry out a numerical study of $P^{x}$
using a hydrodynamic simulation as a cross-check for such an analytical
solution. The prediction of $P^{x}$ can be tested in experiments
and may provide a complete and consistent picture for spin phenomena
in heavy-ion collisions.

The paper is organized as follows. In Sec. \ref{sec:model-description}
we introduce the approach to calculating the spin polarization observables.
In Sec. \ref{sec:method}, we describe the perturbation method based
on the flow-momentum correspondence. In Sec. \ref{sec:general-result},
we present the analytical expressions for the in-plane spin polarization
$P^{x}$, the main results of the paper. Section \ref{sec:numerical}
presents numerical results for analytical solutions and hydrodynamic
simulations. A summary of the paper is given in the final section.


\section{Model description \label{sec:model-description}}

The extended blast wave model provides a unified framework for description
of transverse mass spectra, elliptic flow, and two-particle correlations
\citep{Retiere:2003kf,Jaiswal:2015saa,Yang:2016rnw,Yang:2020oig,Yang:2022yxa}.
It offers a straightforward parameterization of the system at kinetic
freeze-out, characterized by the temperature, transverse flow, and
transverse radius of the source \citep{Siemens:1978pb,Lee:1990sk,Schnedermann:1993ws,Huovinen:2001cy}.

We consider the non-central collision of two high-energy nuclei moving
with the speed of light along the $\pm z$ direction at $x=\pm b/2$.
The direction perpendicular to the reaction plane is then the $y$ direction.

The flow four-velocity and the particle's four-momentum can be parameterized
as
\begin{align}
u^{\mu}(x)= & \left(\cosh\eta\cosh\rho,\sinh\rho\cos\phi_{b},\sinh\rho\sin\phi_{b},\sinh\eta\cosh\rho\right),\label{eq:flow-velocity}\\
p^{\mu}= & \left(m_{T}\cosh Y,p_{T}\cos\phi_{p},p_{T}\sin\phi_{p},m_{T}\sinh Y\right).\label{eq:particle-momentum}
\end{align}
Here $\eta$ and $Y$ are the space-time and momentum rapidity respectively,
$p_{T}$ is the transverse momentum, $m_{T}=\sqrt{m^{2}+p_{T}^{2}}$
is the transverse mass, $\phi_{b}$ is the azimuthal angle in the
transverse flow, the transverse expansion of the fireball \citep{Lee:1990sk,Huovinen:2001cy,Retiere:2003kf}
is described by the transverse rapidity $\rho$ as a function of $r$
(transverse radius), $\phi_{b}$ (azimuthal angle of the flow velocity
in transverse plane), $\epsilon$ (the eccentricity parameter in transverse
emission area), and $\eta$ as follows
\begin{equation}
\rho\left(r,\phi_{b},\eta\right)\approx\frac{r}{R}\left[\rho_{0}+\rho_{1}(\eta)\cos(\phi_{b})+\left(\rho_{2}+\frac{1}{2}\epsilon\rho_{0}\right)\cos(2\phi_{b})\right],\label{eq:rho-r-phi-eta}
\end{equation}
where $\rho_{0}$ characterizes the mean transverse rapidity of the
source element, $\rho_{1}(\eta)=\alpha_{1}\eta$ and $\rho_{2}$ describe
the azimuthal anisotropy of the transverse rapidity. The transverse
emission area can be characterized by $R_{x}$ and $R_{y}$, effective
radii of the elliptic source in $x$- and $y$-direction respectively,
with $R = (R_x + R_y)/2$.
In deriving Eq. (\ref{eq:rho-r-phi-eta}) we used the approximation
$\epsilon\equiv(R_{y}-R_{x})/R\ll1$
which works very well in describing data \citep{Retiere:2003kf}.
We assume 
$\alpha_{1}\sim\rho_{2}\sim\epsilon\ll\rho_{0}$,
so that $\alpha_{1}$, $\rho_{2}$ and $\epsilon$ can be treated as perturbations
relative to $\rho_{0}$. Note that $\phi_{b}$ in the transverse flow
velocity is related to the azimuthal angle $\phi_{s}$ in transverse
coordinate space of the emission source by $\tan\phi_{b}=(R_{x}^{2}/R_{y}^{2})\tan\phi_{s}$,
hence the difference between $\phi_{b}$ and $\phi_{s}$ is $O(\epsilon)$.

The particle's distribution function in phase space is assumed to
follow the Boltzmann distribution, $f(x,p)\equiv f(p\cdot u)=\exp(-\beta p\cdot u)$,
where $\beta=1/T$ is the inverse temperature and $p\cdot u$ is given
by
\begin{equation}
p\cdot u=m_{T}\cosh\rho\cosh(\eta-Y)-p_{T}\sinh\rho\cos(\phi_{b}-\phi_{p}).\label{eq:p-u-contraction}
\end{equation}
We see that $f(p\cdot u)$ depends on rapidities and azimuthal angles
through $\eta-Y$ and $\phi_{b}-\phi_{p}$. This distribution 
reaches a maximum when $\eta\approx Y$ and $\phi_{b}\approx\phi_{p}$,
i.e. the spacetime and momentum rapidities are equal and the flow
and momentum azimuthal angles are equal. If we set $\eta=Y$ and $\phi_{b}=\phi_{p}$,
the distribution reaches a maximum at $p_{T}/m_{T}=\tanh\rho$, meaning
that the transverse momentum rapidity is equal to the transverse flow
rapidity. These equalities between flow variables and particle momentum
variables are called the flow-momentum correspondence in the fireball's
expansion.

The observables can be calculated on the freeze-out hypersurface $\Sigma$
defined by equal temperature condition at the freeze-out proper time
$\tau=\tau_{f}$, $T(\tau_{f},\eta,r,\phi_{s})=T_{f}$, which is a
three-dimensional manifold in $\eta$, $r$ and $\phi_{s}$. On the
freeze-out hypersurface, the partial derivatives of $T$ with respect
to $\eta$, $r$ and $\phi_{s}$ are assumed to vanish except that
with respect to $\tau$, $\left.\partial T/\partial\tau\right|_{\Sigma}\neq0$.
The emission function $S(x,p)$ represents the probability of emitting
a particle with the momentum $p$ at the space-time $x$ incorporated
with the freeze-out condition \citep{Arslan:2024dwi},
\begin{align}
S(x,p) = & m_{T}\cosh(\eta-Y)\delta(\tau-\tau_{f})\Theta(1-\tilde{r})f(p\cdot u)\nonumber\\\approx &  m_{T}\cosh(\eta-Y)\delta(\tau-\tau_{f})\Theta(R-r)f(p\cdot u)\nonumber \\
& -\frac{\epsilon}{2}\cos(2\phi_s)m_{T}\cosh(\eta-Y)\delta(\tau-\tau_{f})\delta(1-r/R)f(p\cdot u),\label{eq:emission-function}
\end{align}
where the last term in the second equality originates from the elliptic deformation of the emission zone
characterized by the normalized elliptic radius \cite{Retiere:2001ed} defined as
\begin{equation}
\tilde{r}=\sqrt{\left(\frac{r}{R_x}\right)^2\cos^2\phi_s+\left(\frac{r}{R_x}\right)^2\sin^2\phi_s} \; .
\end{equation}
Then the expectation value of an observable as a function of three-momentum
(or $p_{T}$, $\phi_{p}$ and $Y$) can be calculated as
\begin{align}
\left\langle O\right\rangle (\mathbf{p})= & \frac{\int d^{4}x\hat{O}(x,p)S(x,p)}{\int d^{4}xS(x,p)}\nonumber \\
= & \frac{\int_{0}^{R}dr\int_{-\infty}^{\infty}d\eta\int_{0}^{2\pi}d\phi_{s}\;rm_{T}\cosh(\eta-Y)\exp(-\beta p\cdot u)\hat{O}(x,p)}{\int_{0}^{R}dr\int_{-\infty}^{\infty}d\eta\int_{0}^{2\pi}d\phi_{s}\;rm_{T}\cosh(\eta-Y)\exp(-\beta p\cdot u)}\nonumber\\
& -\frac{\epsilon R^2 \int_{-\infty}^{+\infty}d\eta\int_{0}^{2\pi}\cos(2\phi_s)d\phi_{s}m_T \cosh(\eta-Y)\exp(-\beta p\cdot u)\hat{O}(x,p)|_{r=R}}{2\int_{0}^{R}dr\int_{-\infty}^{\infty}d\eta\int_{0}^{2\pi}d\phi_{s}\;rm_{T}\cosh(\eta-Y)\exp(-\beta p\cdot u)},\label{eq:weight-ob}
\end{align}
where $\hat{O}(x,p)$ is the observable in phase space and $d^{4}x=\tau rd\tau d\eta drd\phi_{s}$.
The observable's spectra in some variables of $p_{T}$, $\phi_{p}$
and $Y$ can be obtained by integration over the rest of the variables in
both numerator and denominator in Eq. (\ref{eq:weight-ob}). Note that
the observables we consider in this paper are all up to $O(\epsilon)$,
so the $O(\epsilon)$ contribution in the denominators in Eq. (\ref{eq:weight-ob}) can be neglected
since the numerators are already up to $O(\epsilon)$.


\section{Perturbation method and flow-momentum correspondence \label{sec:method}}

From Eqs. (\ref{eq:rho-r-phi-eta},\ref{eq:p-u-contraction},\ref{eq:emission-function}),
the distribution function $f(p\cdot u)$ depends on $\phi_{b}$, so
we have to convert the integral over $\phi_{s}$ in Eq. (\ref{eq:weight-ob})
to that over $\phi_{b}$ by rewriting $\phi_{s}$ in $\hat{O}(x,p)$
in terms of $\phi_{b}$. The integrals over $\eta$ and $\phi_{s}$
(or equivalently $\phi_{b}$) can be carried out using a perturbation method.
To $O(\epsilon)$, the Boltzmann distribution function can be approximated
as
\begin{align}
\exp(-\beta p\cdot u)\approx & \exp\left[-\beta m_{T}\cosh\bar{\rho}\cosh(\Delta\eta)+\beta p_{T}\sinh\bar{\rho}\cos(\Delta\phi)\right]\nonumber \\
 & \times\left[1-\delta\rho\beta m_{T}\sinh\bar{\rho}\cosh(\Delta\eta)+\delta\rho\beta p_{T}\cosh\bar{\rho}\cos(\Delta\phi)\right],\label{eq:exp-exp}
\end{align}
where $\bar{\rho}\equiv(r/R)\rho_{0}$, $\Delta\eta\equiv\eta-Y$,
$\Delta\phi\equiv\phi_{b}-\phi_{p}$ and $\delta\rho$ is given by
\begin{align}
\delta\rho= & \frac{r}{R}\left[\rho_{1}(\eta)\cos(\phi_{b})+\left(\rho_{2}+\frac{1}{2}\epsilon\rho_{0}\right)\cos(2\phi_{b})\right]\nonumber \\
= & \frac{r}{R}\left[\rho_{1}(\Delta\eta+Y)\cos(\Delta\phi)\cos(\phi_{p})-\rho_{1}(\Delta\eta+Y)\sin(\Delta\phi)\sin(\phi_{p})\right.\nonumber \\
 & \left.+\left(\rho_{2}+\frac{1}{2}\epsilon\rho_{0}\right)\cos(2\Delta\phi)\cos(2\phi_{p})-\left(\rho_{2}+\frac{1}{2}\epsilon\rho_{0}\right)\sin(2\Delta\phi)\sin(2\phi_{p})\right],\label{eq:delta-rho}
\end{align}
We see that $\delta\rho$ depends on $\Delta\eta$, $Y$, $\Delta\phi$
and $\phi_{p}$ and is of $O(\epsilon)$. Applying the identity $\tan(\phi_b)=(R_x^2/R_y^2)\tan(\phi_s)$,
the integral measure for $\phi_{s}$ in Eq. (\ref{eq:weight-ob}) can be converted to that
for $\phi_{b}$ to $O(\epsilon)$ as
\begin{align}
d\phi_{s}\approx & d\phi_{b}\left[1+2\epsilon\cos(2\phi_{b})\right]\nonumber \\
= & d\phi_{b}\left[1+2\epsilon\cos(2\Delta\phi)\cos(2\phi_{p})-2\epsilon\sin(2\Delta\phi)\sin(2\phi_{p})\right].\label{eq:dphi-s-phi-b}
\end{align}
We note that the expansion in Eq. (\ref{eq:exp-exp}) is only valid
inside the integral since for large $\bar{\rho}$ and $\Delta\eta$,
$\mathcal{O}(\epsilon)$ terms can easily become larger than 1. When integrated
over $r$ and $\Delta\eta$, the exponential factor suppresses the
contribution from large $\Delta\eta$ ($\cosh\bar{\rho}$ and $\sinh\bar{\rho}$
are finite since $\bar{\rho}<1$ for $\rho_{0}\sim1$).

We now look at the polarization observables $\hat{O}(x,p)=\hat{P}^{i}(x,p)$
with $i=x,y,z$. Normally they depend on $\eta$, $\phi_{s}$ and
$\phi_{b}$ through functions of $\sinh\eta$, $\cosh\eta$, $\sin\phi_{s/b}$
and $\cos\phi_{s/b}$. With $\eta=\Delta\eta+Y$ and $\phi_{b}=\Delta\phi+\phi_{p}$,
the polarization observables can be expressed as functions of $\sinh\Delta\eta$,
$\cosh\Delta\eta$, $\sin\Delta\phi$ and $\cos\Delta\phi$. The integrals
over $\eta$ and $\phi_{s}$ in the numerator in Eq. (\ref{eq:weight-ob})
can be schematically written as
\begin{align}
I_{\eta,\phi}= & \int_{-\infty}^{\infty}d\Delta\eta\int_{0}^{2\pi}d\Delta\phi\;\cosh(\Delta\eta)F\left(r,\Delta\eta,\Delta\phi,p_{T},Y,\phi_{p}\right)\nonumber \\
 & \times\exp\left[-\beta m_{T}\cosh\bar{\rho}\cosh(\Delta\eta)+\beta p_{T}\sinh\bar{\rho}\cos(\Delta\phi)\right],\label{eq:i-eta-phi}
\end{align}
where the integrand function $F$ is defined as
\begin{align}
F\left(r,\Delta\eta,\Delta\phi,p_{T},Y,\phi_{p}\right)\approx & \hat{O}\left(r,\Delta\eta,\Delta\phi,p_{T},Y,\phi_{p}\right)\nonumber \\
 & \times\left[1+2\epsilon\cos(2\Delta\phi)\cos(2\phi_{p})-2\epsilon\sin(2\Delta\phi)\sin(2\phi_{p})\right.\nonumber \\
 & \left.-\delta\rho\beta m_{T}\sinh\bar{\rho}\cosh(\Delta\eta)+\delta\rho\beta p_{T}\cosh\bar{\rho}\cos(\Delta\phi)\right].
\end{align}
The factor inside the square brackets is the product of the second
factor in Eq. (\ref{eq:exp-exp}) and the factor in Eq. (\ref{eq:dphi-s-phi-b}).
Note that in Eq. (\ref{eq:i-eta-phi}) $\bar{\rho}$ only depends
on $r$, so the integral $I_{\eta,\phi}$ can be completed through
formulas for modified Bessel functions of the first and second kinds.

The flow-momentum correspondence in central rapidity means setting
$\eta=Y=0$ in $F\left(r,\Delta\eta,\Delta\phi,p_{T},Y,\phi_{p}\right)$,
so $F$ becomes a function of $r$, $p_{T}$, $Y$, $\phi_{s}$($\phi_{b}$)
and $\phi_{p}$,
\begin{align}
F\left(r,0,\Delta\phi,p_{T},0,\phi_{p}\right)\approx & \hat{O}\left(r,0,\Delta\phi,p_{T},0,\phi_{p}\right)\nonumber \\
 & \times\left[1+2\epsilon\cos(2\Delta\phi)\cos(2\phi_{p})-2\epsilon\sin(2\Delta\phi)\sin(2\phi_{p})\right.\nonumber \\
 & \left.-\left.\delta\rho\right|_{\eta=Y=0}\beta m_{T}\sinh\bar{\rho}+\left.\delta\rho\right|_{\eta=Y=0}\beta p_{T}\cosh\bar{\rho}\cos(\Delta\phi)\right],\label{eq:expansion_F}
\end{align}
where $\left.\delta\rho\right|_{\eta=Y=0}$ is given by
\begin{align}
\left.\delta\rho\right|_{\eta=Y=0} & =\frac{r}{R}\left[\left(\rho_{2}+\frac{1}{2}\epsilon\rho_{0}\right)\cos(2\Delta\phi)\cos(2\phi_{p})\right.\nonumber \\
 & \left.-\left(\rho_{2}+\frac{1}{2}\epsilon\rho_{0}\right)\sin(2\Delta\phi)\sin(2\phi_{p})\right].
\end{align}
Equation (\ref{eq:expansion_F}) is a good approximation at RHIC energy with $\beta p_{T} \delta \rho\ll 1$ being well  satisfied but it is not at LHC energy.
In this paper we will calculate the expectation values of observables
to $O(\epsilon)$ with the flow-momentum correspondence in central
rapidity $\eta=Y=0$ but without imposing that in the azimuthal angle,
i.e. $\Delta\phi\neq0$. This is different from Ref. \citep{Arslan:2024dwi}
in which both $\eta=Y=0$ and $\Delta\phi=0$ were imposed in $F\left(r,\Delta\eta,\Delta\phi,p_{T},Y,\phi_{p}\right)$.


\section{Polarization vectors: general results \label{sec:general-result}}

In this work, we focus on spin-1/2 particles and include polarization
only from the thermal vorticity and thermal shear stress tensors.
There may be other sources of polarizations \citep{Sheng:2021kfc,Banerjee:2024xnd}.
The spin vectors from these two sources are defined as
\begin{align}
\hat{P}_{\omega}^{\mu}= & -\frac{1}{4m}\epsilon^{\mu\nu\sigma\tau}(1-f)\omega_{\nu\sigma}p_{\tau},\label{eq:VIP}\\
\hat{P}_{\xi}^{\mu}= & -\frac{1}{2m}\epsilon^{\mu\nu\sigma\tau}(1-f)\frac{p_{\tau}p^{\rho}}{p\cdot\hat{t}}\hat{t}_{\nu}\xi_{\rho\sigma},\label{eq:SIP}
\end{align}
where the vector $\hat{t}$ is chosen to be $\hat{t}^{\mu}=(1,0,0,0)$
corresponding to the normal direction of the freeze-out hyper-surface $T_f=T(\tau_f)$
for $\eta=Y=0$, $\omega^{\mu\nu}$ and $\xi^{\mu\nu}$ denote the
thermal vorticity and thermal shear stress tensors defined respectively as
\begin{align}
\omega^{\mu\nu}= & -\frac{1}{2}\left[\partial^{\mu}\left(\beta u^{\nu}\right)-\partial^{\nu}\left(\beta u^{\mu}\right)\right]\nonumber \\
= & -\frac{1}{2T}\left(\partial^{\mu}u^{\nu}-\partial^{\nu}u^{\mu}\right)+\frac{1}{2T^{2}}\left(u^{\nu}\partial^{\mu}T-u^{\mu}\partial^{\nu}T\right),\label{eq:vorticity-tensor}\\
\xi^{\mu\nu}= & \frac{1}{2}\left[\partial^{\mu}\left(\beta u^{\nu}\right)+\partial^{\nu}\left(\beta u^{\mu}\right)\right]\nonumber \\
= & \frac{1}{2T}\left(\partial^{\mu}u^{\nu}+\partial^{\nu}u^{\mu}\right)-\frac{1}{2T^{2}}\left(u^{\nu}\partial^{\mu}T+u^{\mu}\partial^{\nu}T\right).\label{eq:shear-tensor}
\end{align}
We see that both $\omega^{\mu\nu}$ and $\xi^{\mu\nu}$ can be decomposed
into kinetic and T-gradient parts.
Note that the definition in Eq.
(\ref{eq:VIP}) is in a Lorentz covariant form and includes the relativistic
effect, different from $\hat{P}_{\omega}^{z}\sim\omega^{xy}$
that was used in Ref. \citep{STAR:2019erd}.

Here we make a few remarks about the freeze-out hyper-surface and the $T$-gradient contribution. In the blast wave model it is normally assumed that $T$ is only a function of the proper time $\tau$, i.e. $T=T(\tau)$, so the $T$-gradient contribution only comes from $\partial T/\partial \tau$. However one can show that there are non-vanishing $\partial T/\partial \tau$ terms in $P_{\omega}$ and $P_{\xi}$, but these terms cancel out in $P_{\omega}+P_{\xi}$ when $\eta =Y=0$ and $\hat{t}^{\mu}=(1,0,0,0)$ (the normal direction of the freeze-out hyper-surface at $\eta=Y=0$).
On the other hand, it can be shown in quantum statistical theory that if $\hat{t}^{\mu}$ is chosen to be the normal direction of the isothermal freeze-out hyper-surface, the $T$-gradient contribution is vanishing to any order of gradient expansion \citep{Sheng:2025cjk}.

With the flow velocity given by Eq. (\ref{eq:flow-velocity}), the
thermal vorticity and thermal shear stress tensors can be evaluated
and give $\hat{P}_{\omega}^{x,y,z}$ as functions of $(\eta,r,\phi_{s},\phi_{b})$
and $(Y,p_{T},\phi_{p})$.
The expectation values of polarization vectors as functions of $\phi_{p}$
can be obtained from Eq. (\ref{eq:weight-ob}) by integration over
$Y$ and $p_{T}$ in both the numerator and denominator as
\begin{align}
P^{x}(\phi_{p})= & \left\langle \hat{P}^{x}\right\rangle (\phi_{p})=-\alpha_{1}\frac{1}{16mT\tau R}\sin(2\phi_{p})\nonumber \\
 & \times\frac{1}{N_{0}}\left[N_{c}(0|2,3,0)+N_{c}(2|2,3,0)+2N_{c}(2|2,1,2)-4N_{s}(1|2,2,1)\right],\nonumber \\
P^{y}(\phi_{p})= & \left\langle \hat{P}^{y}\right\rangle (\phi_{p})\nonumber \\
= & \alpha_{1}\frac{1}{8mT\tau R}\frac{1}{N_{0}}\left[N_{c}(0|2,1,2)-N_{c}(2|2,1,2)\right]+\alpha_{1}\frac{1}{8mT\tau R}\cos^{2}\phi_{p}\nonumber \\
 & \times\frac{1}{N_{0}}\left[N_{c}(0|2,3,0)+N_{c}(2|2,3,0)+2N_{c}(2|2,1,2)-4N_{s}(1|2,2,1)\right],\nonumber \\
P^{z}(\phi_{p})= & \left\langle \hat{P}^{z}\right\rangle (\phi_{p})\nonumber \\
= & \frac{1}{8mT}\sin(2\phi_{p})\frac{1}{N_{0}}\int_{0}^{p_{T}^{\mathrm{max}}}dp_{T}\int_{0}^{R}dr\;rp_{T}\nonumber \\
 & \times\sum_{n=0}\left(m_{T}C_{z,n}^{\omega}+p_{T}C_{z,n}^{\xi}\right)K_{1}(\beta m_{T}\cosh\bar{\rho})I_{n}(\beta p_{T}\sinh\bar{\rho})\nonumber\\&+\frac{\epsilon R}{16mT}\sin(2\phi_{p})\frac{1}{N_{0}}\int_{0}^{p_{T}^{\mathrm{max}}}dp_{T}K_{1}\left(\beta m_{T}\cosh\rho_{0}\right)\nonumber\\&\times\left\{ p_{T}^{3}\left(\sinh\rho_{0}-\rho_{0}\cosh\rho_{0}\right)\right.\left[I_{0}\left(\beta p_{T}\sinh\rho_{0}\right)-I_{4}\left(\beta p_{T}\sinh\rho_{0}\right)\right]\nonumber\\&\left.+2p_{T}^{2}m_{T}\rho_{0}\sinh\rho_{0}\left[I_{1}\left(\beta p_{T}\sinh\rho_{0}\right)-I_{3}\left(\beta p_{T}\sinh\rho_{0}\right)\right]\right\} ,\label{eq:pol-xyz-phi-p}
\end{align}
where $\hat{P}^{i}=\hat{P}_{\omega}^{i}+\hat{P}_{\xi}^{i}$ with $i=x,y,z$
and
\begin{align}
N_{0}= & \int_{0}^{p_{T}^{\mathrm{max}}}dp_{T}\int_{0}^{R}dr\;rp_{T}m_{T}\;K_{1}(\beta m_{T}\cosh\bar{\rho})I_{0}(\beta p_{T}\sinh\bar{\rho}),\nonumber \\
N_{c}(n|n_{1},n_{2},n_{3})= & \int_{0}^{p_{T}^{\mathrm{max}}}dp_{T}\int_{0}^{R}dr\;r^{n_{1}}p_{T}^{n_{2}}m_{T}^{n_{3}}\;\cosh\bar{\rho}\;K_{1}(\beta m_{T}\cosh\bar{\rho})I_{n}(\beta p_{T}\sinh\bar{\rho}),\nonumber \\
N_{s}(n|n_{1},n_{2},n_{3})= & \int_{0}^{p_{T}^{\mathrm{max}}}dp_{T}\int_{0}^{R}dr\;r^{n_{1}}p_{T}^{n_{2}}m_{T}^{n_{3}}\;\sinh\bar{\rho}\;K_{1}(\beta m_{T}\cosh\bar{\rho})I_{n}(\beta p_{T}\sinh\bar{\rho}).\label{eq:ni-pt-int}
\end{align}
The explicit forms of $C^\omega_{z,n}$ and $C^{\xi}_{z,n}$ in Eq. (\ref{eq:pol-xyz-phi-p})
are given in Appendix \ref{appendix-A}.

\begin{figure}
\begin{centering}
\includegraphics[scale=0.8]{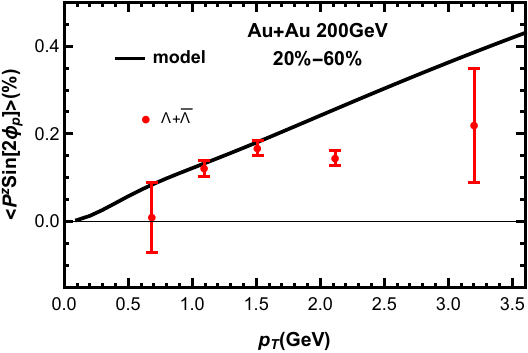} \includegraphics[scale=0.8]{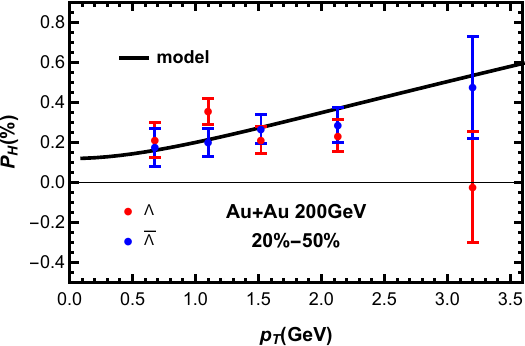}
\par\end{centering}
\caption{The calculated results for $\left\langle P^{z}\sin(2\phi_{p})\right\rangle $
and $P_{H}$ as functions of $p_{T}$ from Eq. (\ref{eq:pol-xyz-pt}).
The polarization at 20-60\% and 20-50\% is calculated by the total particle-production weighted average.}
\label{fig:P_pT}
\end{figure}

Similarly the integrated polarization vectors can be obtained by further
integrating over $\phi_{p}$ from Eq. (\ref{eq:pol-xyz-phi-p}). But
$P^{x}(\phi_{p})$ and $P^{z}(\phi_{p})$ are proportional to $\sin(2\phi_{p})$
whose integration over $\phi_{p}$ is vanishing. In order to obtain
meaningful results for integrated polarization vectors in $x$ and
$z$ directions, we can calculate weighted observables $P_{\sin2\phi}^{x}=\left\langle \hat{P}^{x}\sin(2\phi_{p})\right\rangle $
and $P_{\sin2\phi}^{z}=\left\langle \hat{P}^{z}\sin(2\phi_{p})\right\rangle $.
For the longitudinal polarization, we can calculate $P^{y}=\left\langle \hat{P}^{y}\right\rangle $
or $P_{\cos2\phi}^{y}=\left\langle \hat{P}^{y}\cos(2\phi_{p})\right\rangle $.
The results are
\begin{align}
P_{\sin2\phi}^{x}= & \left\langle \hat{P}^{x}\sin(2\phi_{p})\right\rangle =-\alpha_{1}\frac{1}{32mT\tau R}\nonumber \\
 & \times\frac{1}{N_{0}}\left[N_{c}(0|2,3,0)+N_{c}(2|2,3,0)+2N_{c}(2|2,1,2)-4N_{s}(1|2,2,1)\right],\nonumber \\
P^{y}= & \left\langle \hat{P}^{y}\right\rangle =\alpha_{1}\frac{1}{16mT\tau R}\nonumber \\
 & \times\frac{1}{N_{0}}\left[N_{c}(0|2,3,0)+N_{c}(2|2,3,0)+2N_{c}(0|2,1,2)-4N_{s}(1|2,2,1)\right],\nonumber \\
P_{\cos2\phi}^{y}= & \left\langle \hat{P}^{y}\cos(2\phi_{p})\right\rangle =\alpha_{1}\frac{1}{32mT\tau R}\nonumber \\
 & \times\frac{1}{N_{0}}\left[N_{c}(0|2,3,0)+N_{c}(2|2,3,0)+2N_{c}(2|2,1,2)-4N_{s}(1|2,2,1)\right],\nonumber \\
P_{\sin2\phi}^{z}= & \left\langle \hat{P}^{z}\sin(2\phi_{p})\right\rangle \nonumber \\
= & \frac{1}{16mT}\frac{1}{N_{0}}\int_{0}^{p_{T}^{\mathrm{max}}}dp_{T}\int_{0}^{R}dr\;rp_{T}\nonumber \\
 & \times\sum_{n=0}\left(m_{T}C_{z,n}^{\omega}+p_{T}C_{z,n}^{\xi}\right)K_{1}(\beta m_{T}\cosh\bar{\rho})I_{n}(\beta p_{T}\sinh\bar{\rho})\nonumber\\&+\frac{\epsilon R}{32mT}\frac{1}{N_{0}}\int_{0}^{p_{T}^{\mathrm{max}}}dp_{T}K_{1}\left(\beta m_{T}\cosh\rho_{0}\right)\nonumber\\&\times\left\{ p_{T}^{3}\left(\sinh\rho_{0}-\rho_{0}\cosh\rho_{0}\right)\right.\left[I_{0}\left(\beta p_{T}\sinh\rho_{0}\right)-I_{4}\left(\beta p_{T}\sinh\rho_{0}\right)\right]\nonumber\\&\left.+2p_{T}^{2}m_{T}\rho_{0}\sinh\rho_{0}\left[I_{1}\left(\beta p_{T}\sinh\rho_{0}\right)-I_{3}\left(\beta p_{T}\sinh\rho_{0}\right)\right]\right\}.\label{eq:integrated-pol}
\end{align}
The only difference between $P^{y}$ and $P_{\cos2\phi}^{y}$ is that
the former has a term $2N_{c}(0|2,1,2)$ while the latter has a term
$2N_{c}(2|2,1,2)$. Apart from such a difference, we observe an approximated equality
for $P^{y}$ and $P_{\sin2\phi}^{x}$: $P^{y}\approx -2P_{\sin2\phi}^{x}$,
which will be confirmed by numerical results in Figs. \ref{fig:P_pT}, \ref{fig:Pz_centrality} and \ref{fig:pol-x}.
Furthermore, there is an exact equality for $P_{\sin2\phi}^{x}$ and $P_{\cos2\phi}^{y}$: $P_{\sin2\phi}^{x}=-P_{\cos2\phi}^{y}$, showing a good symmetry between the in-plane and out-of-plane polarization.

In a similar way, the expectation values of polarization or weighted
polarization vectors as functions of $p_{T}$ can be obtained from
Eq. (\ref{eq:weight-ob}) by integration over $Y$ and $\phi_{p}$
in both the numerator and denominator as
\begin{align}
P_{\sin2\phi}^{x}(p_{T})= & \left\langle \hat{P}^{x}\sin(2\phi_{p})\right\rangle (p_{T})=-\alpha_{1}\frac{1}{32mT\tau R}\nonumber \\
 & \times\frac{1}{N_{0}(p_{T})}\left[N_{c}^{p}(0|2,2,0)+N_{c}^{p}(2|2,2,0)+2N_{c}^{p}(2|2,0,2)-4N_{s}^{p}(1|2,1,1)\right],\nonumber \\
P^{y}(p_{T})= & \left\langle \hat{P}^{y}\right\rangle (p_{T})=\alpha_{1}\frac{1}{16mT\tau R}\nonumber \\
 & \times\frac{1}{N_{0}(p_{T})}\left[N_{c}^{p}(0|2,2,0)+N_{c}^{p}(2|2,2,0)+2N_{c}^{p}(0|2,0,2)-4N_{s}^{p}(1|2,1,1)\right],\nonumber \\
P_{\cos2\phi}^{y}(p_{T})= & \left\langle \hat{P}^{y}\cos(2\phi_{p})\right\rangle (p_{T})=\alpha_{1}\frac{1}{32mT\tau R}\frac{1}{N_{0}(p_{T})}\nonumber \\
 & \times\left[N_{c}^{p}(0|2,2,0)+N_{c}^{p}(2|2,2,0)+2N_{c}^{p}(2|2,0,2)-4N_{s}^{p}(1|2,1,1)\right],\nonumber \\
P_{\sin2\phi}^{z}(p_{T})= & \left\langle \hat{P}^{z}\sin(2\phi_{p})\right\rangle (p_{T})\nonumber \\
= & \frac{1}{16mT}\frac{1}{N_{0}(p_{T})}\int_{0}^{R}dr\;r\sum_{n=0}\left(m_{T}C_{z,n}^{\omega}+p_{T}C_{z,n}^{\xi}\right)\nonumber \\
 & \times K_{1}(\beta m_{T}\cosh\bar{\rho})I_{n}(\beta p_{T}\sinh\bar{\rho})\nonumber\\&+\frac{\epsilon R}{32mT}\frac{1}{N_{0}(p_T)}K_{1}\left(\beta m_{T}\cosh\rho_{0}\right)\nonumber\\&\times\left\{ p_{T}^{2}\left(\sinh\rho_{0}-\rho_{0}\cosh\rho_{0}\right)\right.\left[I_{0}\left(\beta p_{T}\sinh\rho_{0}\right)-I_{4}\left(\beta p_{T}\sinh\rho_{0}\right)\right]\nonumber\\&\left.+2p_{T}m_{T}\rho_{0}\sinh\rho_{0}\left[I_{1}\left(\beta p_{T}\sinh\rho_{0}\right)-I_{3}\left(\beta p_{T}\sinh\rho_{0}\right)\right]\right\},\label{eq:pol-xyz-pt}
\end{align}
where $N_{0}(p_{T})$ and $N_{i}^{p}(n|n_{1},n_{2},n_{3})$ ($i=c,s$)
are functions of $p_{T}$ defined as
\begin{align}
N_{0}(p_{T})= & \int_{0}^{R}dr\;rm_{T}\;K_{1}(\beta m_{T}\cosh\bar{\rho})I_{0}(\beta p_{T}\sinh\bar{\rho}),\nonumber \\
N_{c}^{p}(n|n_{1},n_{2},n_{3})= & \int_{0}^{R}dr\;r^{n_{1}}p_{T}^{n_{2}}m_{T}^{n_{3}}\;\cosh\bar{\rho}\;K_{1}(\beta m_{T}\cosh\bar{\rho})I_{n}(\beta p_{T}\sinh\bar{\rho}),\nonumber \\
N_{s}^{p}(n|n_{1},n_{2},n_{3})= & \int_{0}^{R}dr\;r^{n_{1}}p_{T}^{n_{2}}m_{T}^{n_{3}}\;\sinh\bar{\rho}\;K_{1}(\beta m_{T}\cosh\bar{\rho})I_{n}(\beta p_{T}\sinh\bar{\rho}).\label{eq:ni-pt}
\end{align}
We observe that $N_{0}$ and $N_{i}(n|n_{1},n_{2},n_{3})$ ($i=c,s$)
in (\ref{eq:ni-pt-int}) can be obtained by integration of $N_{0}(p_{T})$
and $N_{i}^{p}(n|n_{1},n_{2},n_{3})$ over $p_{T}$ as
\begin{align}
N_{0}= & \int_{0}^{p_{T}^{\mathrm{max}}}dp_{T}p_{T}N_{0}(p_{T}),\nonumber \\
N_{i}(n|n_{1},n_{2}+1,n_{3})= & \int_{0}^{p_{T}^{\mathrm{max}}}dp_{T}p_{T}N_{i}^{p}(n|n_{1},n_{2},n_{3}).
\end{align}
We also observe an approximated equality for $P^{y}(p_{T})$ and $P_{\sin2\phi}^{x}(p_{T})$
from Eq. (\ref{eq:pol-xyz-pt}): $P^{y}\approx -2P_{\sin2\phi}^{x}$,
which will be confirmed by numerical results in Figs. \ref{fig:P_pT}, \ref{fig:Pz_centrality} and \ref{fig:pol-x}.
There is also an exact equality for $P_{\sin2\phi}^{x}(p_{T})$ and $P_{\cos2\phi}^{y}(p_{T})$: $P_{\sin2\phi}^{x}(p_{T})=-P_{\cos2\phi}^{y}(p_{T})$, which shows a good symmetry between the in-plane and out-of-plane polarization.


\section{Transverse polarizations \label{sec:transverse-polar}}

\begin{figure}
\includegraphics[scale=0.6]{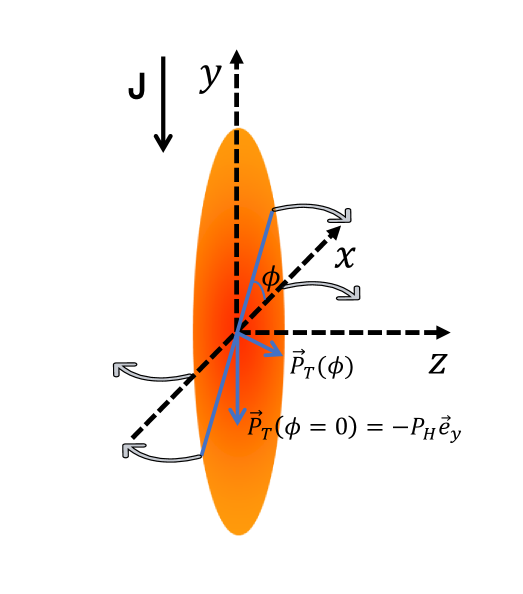}\includegraphics[scale=0.6]{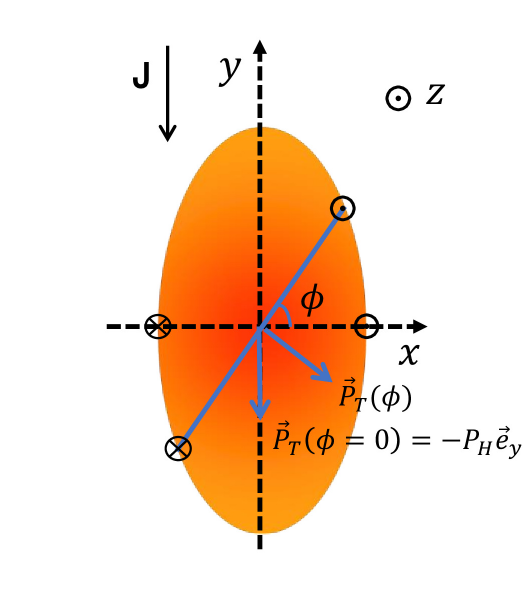}

\caption{The transverse spin polarization vector $\mathbf{P}_{T}(\phi_{p})$
in Eq. (\ref{eq:transverse-spin}). \label{fig:transverse-spin}}
\end{figure}

\begin{figure}
\includegraphics[scale=0.6]{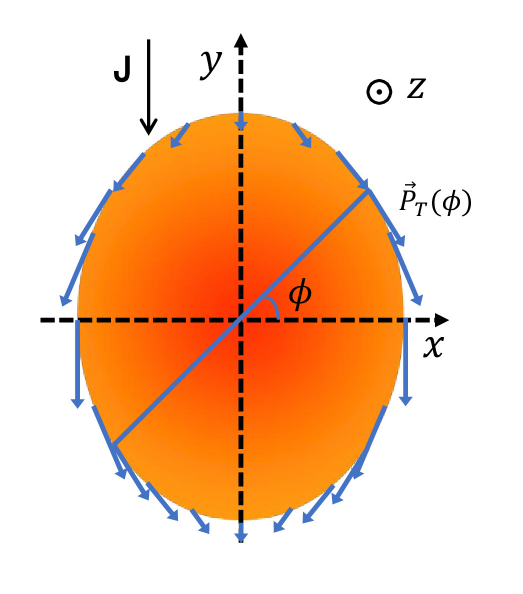}

\caption{\label{fig:vector-plot-pol-t}Vector plot of $\mathbf{P}_{T}(\phi_{p})$
with arrow and length on a circle varying with $\phi_{p}$. }
\end{figure}

We see from Eqs. (\ref{eq:pol-xyz-phi-p},\ref{eq:integrated-pol},\ref{eq:pol-xyz-pt})
that both $P^{x}$ and $P^{y}$ are driven by the directed flow. The
$\phi_{p}$ pattern of $P^{x}$ is $\sin(2\phi_{p})$, same as $P^{z}$
but with a slightly smaller magnitude which will be shown in Fig.
\ref{fig:pol-x}. There are two terms in $P^{y}(\phi_{p})$, the first
term would be vanishing if we take $\Delta\phi=0$ in $\hat{P}^{y}(\phi_{p})$
before the integration over $\Delta\phi$ in calculating its expectation
value as in Ref. \citep{Arslan:2024dwi}, while the second term is
proportional to $P^{x}(\phi_{p})$ up to a modulation factor $\sim\sin\phi_{p}$.
We can combine $P^{x}(\phi_{p})$ and $P^{y}(\phi_{p})$ to form a
vector in the transverse plane
\begin{align}
\mathbf{P}_{T}(\phi_{p})= & \mathbf{e}_{x}P^{x}(\phi_{p})+\mathbf{e}_{y}P^{y}(\phi_{p})\nonumber \\
= & \mathbf{e}_{\phi}\alpha_{1}\frac{1}{8mT\tau R}\cos\phi_{p}\frac{1}{N_{0}}\left[N_{c}(0|2,3,0)+N_{c}(2|2,3,0)\right.\nonumber \\
 & \left.+N_{c}(0|2,1,2)+N_{c}(2|2,1,2)-4N_{s}(1|2,2,1)\right]\nonumber \\
 & +\mathbf{e}_{r}\alpha_{1}\frac{1}{8mT\tau R}\sin\phi_{p}\frac{1}{N_{0}}\left[N_{c}(0|2,1,2)-N_{c}(2|2,1,2)\right]\label{eq:transverse-spin}
\end{align}
where we have used $\mathbf{e}_{\phi}=-\mathbf{e}_{x}\sin\phi_{p}+\mathbf{e}_{y}\cos\phi_{p}$
and $\mathbf{e}_{y}=\mathbf{e}_{r}\sin\phi_{p}+\mathbf{e}_{\phi}\cos\phi_{p}$.
The geometric configuration for $\mathbf{P}_{T}(\phi_{p})$
is shown in Fig. \ref{fig:transverse-spin}. The vector of $\mathbf{P}_{T}(\phi_{p})$
with arrow and length is plotted in Fig. \ref{fig:vector-plot-pol-t}
along an ellipse varying with $\phi_{p}$ (note that $\alpha_{1}<0$).

\section{Numerical results \label{sec:numerical}}

In this section, we will give numerical results for polarization observables
using analytical formulas derived in Sec. \ref{sec:general-result}.
A systematic comparison between numerical results for $P^{y}$ and
$P^{z}$ with experimental data will be performed. We will also give
numerical results for the in-plane polarization $P^{x}$ by hydrodynamical
simulation and compare them with the prediction based on analytical
formulas.

\subsection{Numerical results from analytical formulas}

\begin{figure}
\begin{centering}
\includegraphics[scale=0.42]{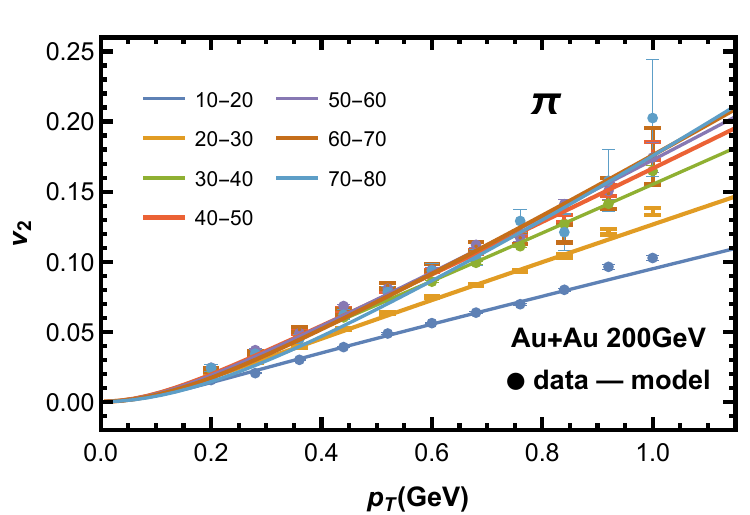} \includegraphics[scale=0.42]{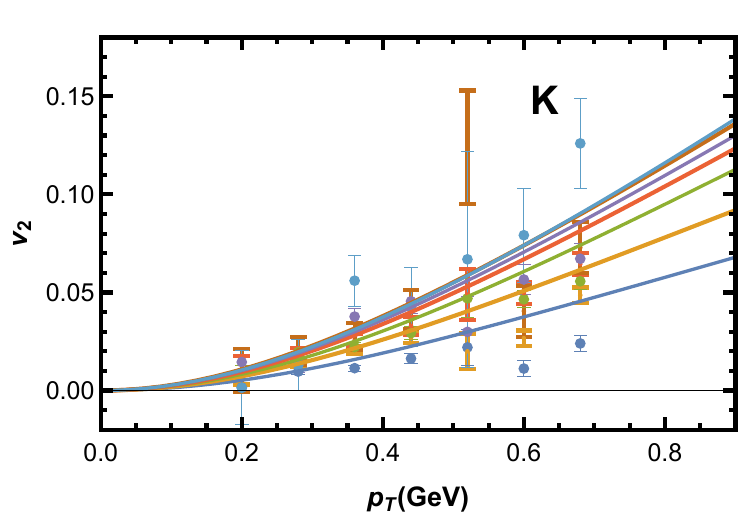} \includegraphics[scale=0.42]{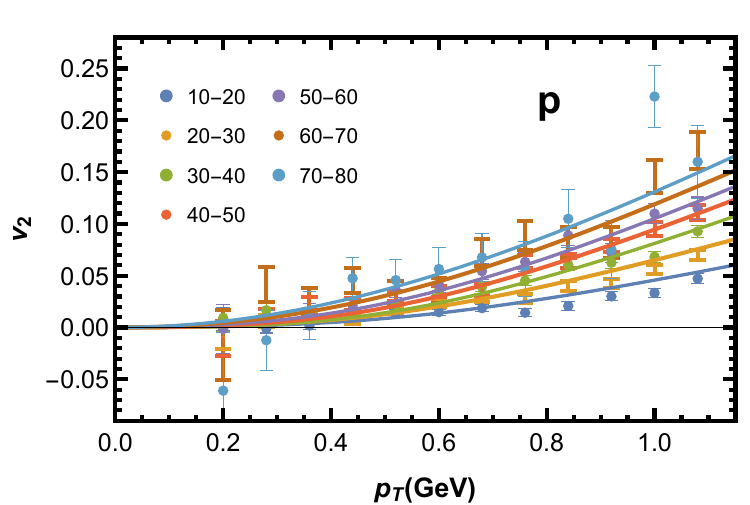}
\par\end{centering}
\caption{The model fit to elliptic flow data of light particles in 10-80\% central Au+Au collisions at 200 GeV with the values of $\rho_2$ and $\epsilon$ listed in Table \ref{tab:centrality-para}. \label{fig:v2}}
\end{figure}

The parameters we choose for Au+Au collisions at $\sqrt{s_{NN}}=200$
GeV and different centralities are listed in Table \ref{tab:centrality-para}.
The main difference in the values of parameters between this paper
and Ref. \citep{Arslan:2024dwi} is in $\rho_{2}$ and $\epsilon$
which are determined by fitting the directed and elliptic flow data.
In Ref. \citep{Arslan:2024dwi}, different combinations of $\rho_{2}$ and $\epsilon$ can fit the
flow data equally well,
but the new values of $\rho_{2}$ and $\epsilon$
in this paper are better suited for the current analytical formulation
of the polarization. In Fig. \ref{fig:v2} we show the model fit to elliptic flow data of light particles in 10-80\% central Au+Au collisions with the values of $\rho_2$ and $\epsilon$ listed in Table \ref{tab:centrality-para}.
The freeze-out temperature $T_{f}$ and transverse
rapidity parameter $\rho_{0}$ are extracted by fitting transverse
momentum spectra \citep{STAR:2004jwm,STAR:2008med}. The parameter
$\alpha_{1}$ is set to $-0.05$ by fitting the data for directed
flows of $\Lambda/\overline{\Lambda}$ in 10-40\% central Au+Au collisions
\citep{STAR:2017okv}. For elliptic flows, we fit the data for light
particles and $\Lambda+\overline{\Lambda}$ in different centralities
\citep{STAR:2004jwm,STAR:2008ftz}. Our results for light particles
in 10-80\% central collisions \citep{STAR:2004jwm} and $\Lambda+\overline{\Lambda}$
in 10-40\% central collisions \citep{STAR:2008ftz} agree well with
experimental data.

\begin{table}
\begin{centering}
\begin{tabular}{|c|c|c|c|c|c|c|c|}
\hline
centrality & $R$ (fm) & $T$ (MeV) & $\rho_{0}$ & $\rho_{2}$ & $\epsilon$ & $\alpha_{1}$ & $\tau_{f}$(fm/c)\tabularnewline
\hline
\hline
10\%-20\% & 11.5 & 99.5 & 0.982 & 0.05 & 0.097 & -0.05 & 7.8\tabularnewline
\hline
20\%-30\% & 10.3 & 102 & 0.937 & 0.068 & 0.128 & -0.05 & 6.9\tabularnewline
\hline
30\%-40\% & 9 & 104 & 0.894 & 0.089 & 0.142 & -0.05 & 5.1\tabularnewline
\hline
40\%-50\% & 7.8 & 107 & 0.841 & 0.097 & 0.154 & -0.05 & 3.3\tabularnewline
\hline
50\%-60\% & 7 & 110 & 0.788 & 0.103 & 0.16 & -0.05 & 2.6\tabularnewline
\hline
60\%-70\% & 6.3 & 116 & 0.707 & 0.11 & 0.173 & -0.05 & 2.3\tabularnewline
\hline
70\%-80\% & 5.5 & 125 & 0.608 & 0.124 & 0.18 & -0.05 & 2\tabularnewline
\hline
\end{tabular}
\par\end{centering}
\caption{Parameters used in this paper for Au+Au collisions at $\sqrt{s_{NN}}=200$
GeV. \label{tab:centrality-para}}
\end{table}

\begin{figure}
\begin{centering}
\includegraphics[scale=0.8]{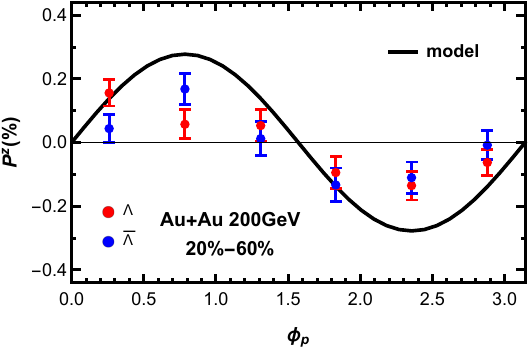} \includegraphics[scale=0.8]{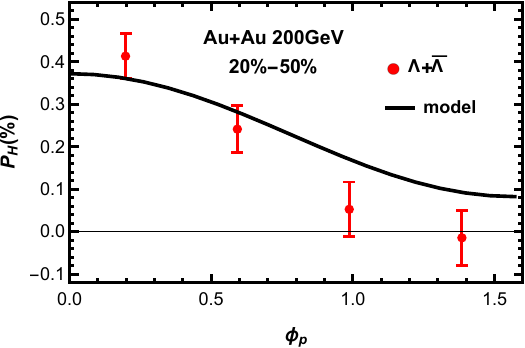}
\par\end{centering}
\caption{The results for $P^{z}$ (left panel) and $P_{H}\equiv-P^{y}$ (right
panel) as functions of $\phi_{p}$ following Eq.(\ref{eq:integrated-pol}).
The black solid lines represent the calculated results based on analytical
formulas. \label{fig:P_phip}}
\end{figure}

The spin polarizations $P^{z}$ and $P_{H}\equiv-P^{y}$ as functions
of $\phi_{p}$ are calculated by Eq. (\ref{eq:pol-xyz-phi-p}). The
experimental data for $P^{z}$ are available 
for the 20-60\% centrality Au+Au collisions at $\sqrt{s_{NN}}=200\text{ GeV}$ \citep{STAR:2019erd},
while the $P^{y}$ data are available 
for the 20-50\% centrality class \citep{STAR:2018gyt}.
The comparison between the calculated results and experimental data
are shown in Fig. \ref{fig:P_phip}. These results are different from
Ref. \citep{Voloshin:2017kqp} in which a non-relativistic approximation
$P^{z}\approx\omega^{z}/2$ was used.

\begin{figure}
\begin{centering}
\includegraphics[scale=0.8]{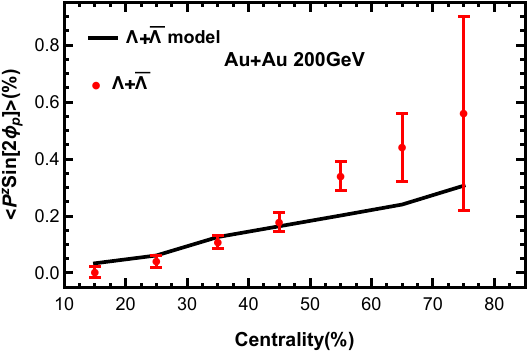} \includegraphics[scale=0.8]{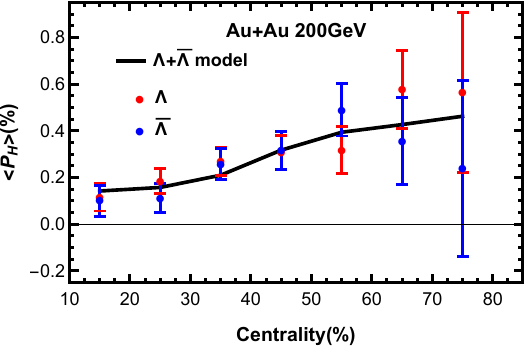}
\par\end{centering}
\caption{The centrality dependence of $P_{z}$ and $P_{H}$.
\label{fig:Pz_centrality}}
\end{figure}

The calculated results for the transverse momentum dependence of polarization
following the analytical formulas in Eq. (\ref{eq:pol-xyz-pt}). In contrast to the analytical formula in \cite{Arslan:2024dwi}, Eq. (\ref{eq:pol-xyz-pt})  fits better to the data.

We can also calculate the centrality dependence of $P^{z}$ in the
form of $\left\langle \hat{P}^{z}\sin(2\phi_{p})\right\rangle $ and
$P_{H}$ using the analytical formulas in Eq. (\ref{eq:integrated-pol}).
The comparison of theoretical results with data is shown in Fig. \ref{fig:Pz_centrality}.
One can see that theoretical curves grow from central to peripheral
collisions which can describe the experimental data.

\begin{figure}
\includegraphics[scale=0.6]{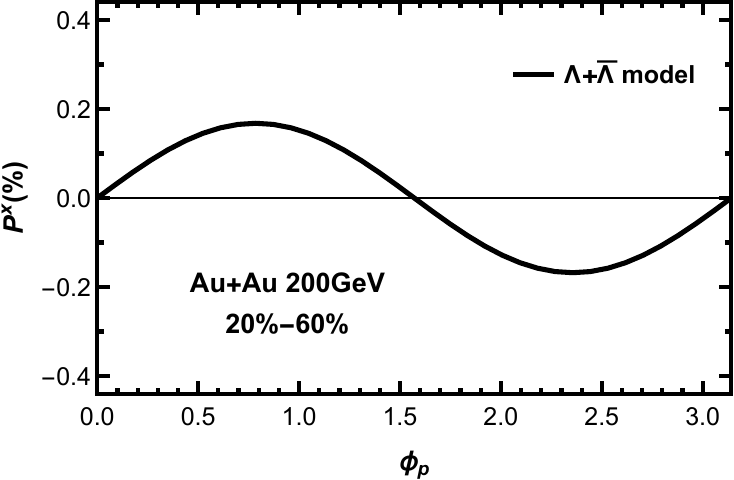}\hspace{0.5cm}\includegraphics[scale=0.6]{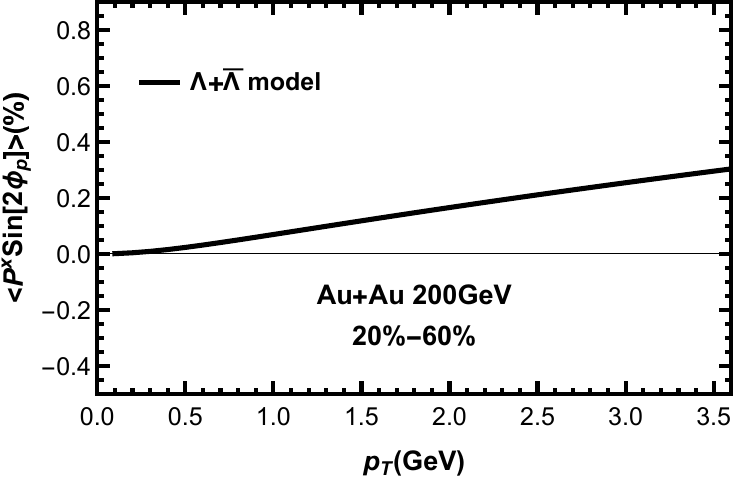}\\  \vspace{0.5cm}\includegraphics[scale=0.6]{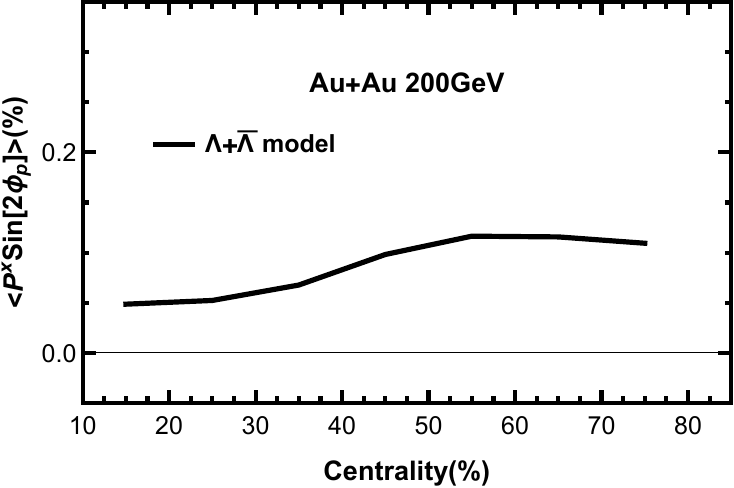}

\caption{The first panel: the in-plane polarization $P^{x}$ as functions of $\phi_{p}$;
The second and third panels: $\left\langle \hat{P}^{x}\sin(2\phi_{p})\right\rangle $ as functions
of $p_{T}$ and centrality.
\label{fig:pol-x}}
\end{figure}

With the values of parameters in Table \ref{tab:centrality-para}
fixed by experimental data for collective flows and polarizations
along the longitudinal and global orbital angular momentum directions,
$P^{z}$ and $P^{y}$, we can make prediction for a new observable,
the in-plane polarization $P^{x}$. The numerical results from analytical
formula are presented in Fig. \ref{fig:pol-x}. We see that the behavior
and magnitude of $P^{x}$ are similar to $P^{z}$ although the former
is driven by the directed flow while the latter is driven by the elliptic
flow.

\subsection{Simulation results by hydrodynamical models}

To compare and understand the polarization results from the blast
wave model, we also calculate the corresponding polarization observables
using the realistic (3+1)D iEBE-MUSIC framework~\cite{Schenke:2010nt,Schenke:2010rr,Paquet:2015lta,Denicol:2018wdp}. This framework includes
smooth initial conditions 
from SMASH, realistic viscous hydrodynamic
evolution, the iSS sampler, and the SMASH  afterburner. Specifically,
for smooth initial conditions 
from SMASH~\cite{SMASH:2016zqf,Schafer:2021csj}, we use a Gaussian smearing
function to construct the initial energy-momentum tensor $T_{0}^{\mu\nu}$
and net baryon current $J_{0}^{\mu}$ when the initial hadrons reach
the hypersurface at the initial proper time $\tau_{0}=$0.5 fm. The
Gaussian widths are $\sigma_{r}=$1.0 fm and $\sigma_{\eta_{s}}=$0.8.
Each smooth initial condition is generated by averaging over 5000
events in each 10\% centrality bin.

The subsequent hydrodynamic evolution included
the net baryon current without diffusion.
The value of the shear viscosity is set to $\eta/s = 0.08$ and the bulk pressure is neglected. The NEOS-BQS equation of state~\cite{Monnai:2019hkn} is
used during the hydrodynamic evolution. When the energy density of
the QGP medium drops to the freeze-out energy density 0.4 GeV/fm$^{3}$,
the QGP converts to soft thermal hadrons via the Cooper-Frye prescription~\cite{Cooper:1974mv}
and the iSS sampler module. Finally, these soft hadrons are injected
into SMASH for further scattering.

For the polarization part, based on the local thermal equilibrium
assumption, the polarization vector for spin-1/2 particles can be
calculated with the modified Cooper-Frye formula~\citep{Becattini:2013fla,Fang:2016vpj}:
\begin{equation}
P^{\mu}=\frac{\int d\Sigma^{\alpha}p_{\alpha}f\hat{P}^{\mu}}{\int d\Sigma^{\alpha}p_{\alpha}f}.\label{eq:pol_ave}
\end{equation}
Here, $f$ is the Fermi-Dirac distribution, and $d\Sigma^{\alpha}$
represents the freeze-out hypersurface elements, determined by the
Cornelius routine. The polarization vector $\hat{P}^{\mu}$ can be
written as:
\begin{equation}
\hat{P}^{\mu}=\hat{P}_{\omega,\mathrm{kin}}^{\mu}+\hat{P}_{\xi,\mathrm{kin}}^{\mu}+\hat{P}_{T}^{\mu}.
\end{equation}
The first two terms 
are the velocity-gradient terms in Eq. (\ref{eq:VIP}) and Eq. (\ref{eq:SIP}) with $\hat{t}^{\mu}=u^{\mu}$.
The last term,
$\hat{P}_{T}^{\mu}$, represents the polarization caused by the temperature
gradient ($T$-gradient) and is given by:
\begin{equation}
\hat{P}_{T}^{\mu}=-\frac{1}{2m}\epsilon^{\mu\nu\sigma\tau}(1-f)\omega_{\nu\sigma}^{T}p_{\tau},
\end{equation}
where the $T$-gradient vorticity is defined as
\begin{equation}
\omega_{\mu\nu}^{T}=\frac{1}{2T^{2}}(u^{\nu}\partial^{\mu}T-u^{\mu}\partial^{\nu}T).
\end{equation}

There are several differences between the hydrodynamic simulation
and the blast wave model.
On the one hand, the calculation of the polarization
in hydrodynamics does not assume an isothermal condition. Although
the temperature on the freeze-out hypersurface is almost constant
if we ignore net baryon density, this does not mean the temperature
gradient on the hypersurface disappears.
For instance, numerical determination of
the freeze-out hypersurface in hydrodynamic simulation relies on temperature gradient which provides a normal vector to the freeze-out hypersurface
\cite{Huovinen:2012is}.
Therefore, the polarization in hydrodynamic simulation includes the
the temperature-gradient contribution. On the other hand, the freeze-out
hypersurface in Eq. (\ref{eq:pol_ave}) is not the same as
the freeze-out hypersurface in the
blast wave model, which only considers time-like terms (such as $\tau dxdyd\eta$)
or the iso-proper-time freeze-out. In hydrodynamic simulation, the
hypersurface also contains spatial components and depends on the evolution
of the QGP medium.
Consequently, all types of vorticity in hydrodynamic
simulation depend on the evolution time $\tau$, while in the blast
wave model, only the vorticity at the final moment is considered.

In order to make a more realistic prediction of the in-plane
transverse polarization $P^{x}$, we consider the s-quark equilibrium
scenario, where the s-quark mass is set to $m_{s}=$0.5 GeV. We also
set the $\hat{t}$-vector in $\hat{P}_{\xi}^{\mu}$ in Eq. (\ref{eq:SIP})
as the fluid four-velocity $u^{\mu}$ in our calculation.

\begin{figure}
\includegraphics[scale=0.5]{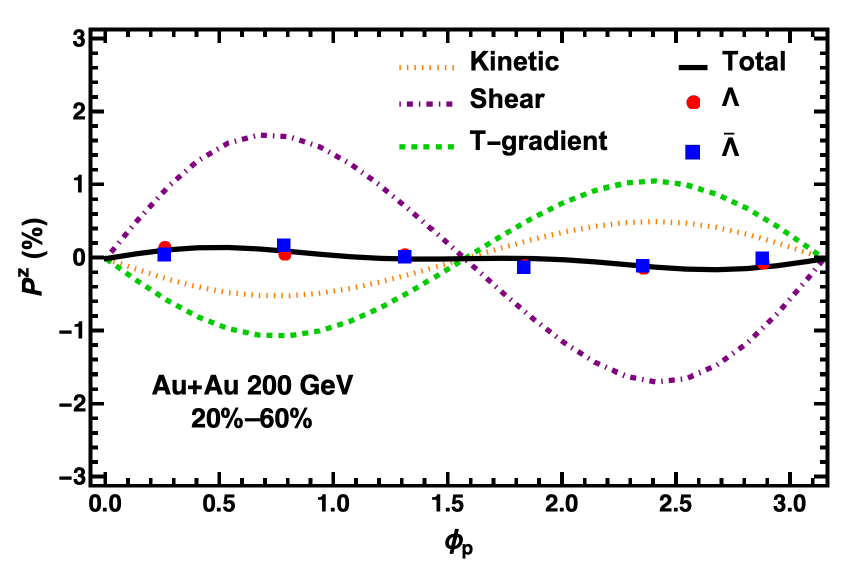}\includegraphics[scale=0.5]{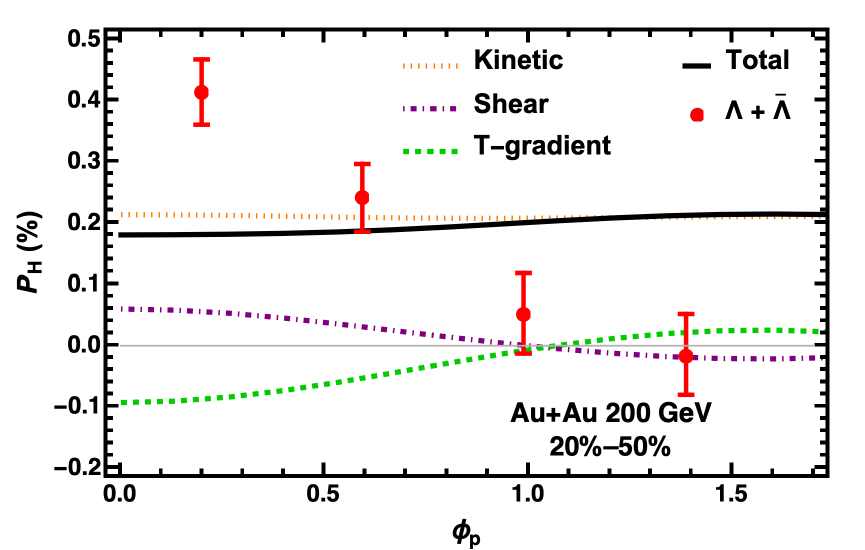}
\caption{The azimuthal angle dependence of the longitudinal polarization
$P^{z}$ (left panel) and the out-of-plane polarization $P_{H}=-P^y$ (right
panel), obtained from hydrodynamic simulation including the kinetic
vorticity, shear vorticity, and temperature-gradient contributions
in Au+Au collisions at 200 GeV. Experimental
data are taken from the STAR Collaboration. \label{fig:pz_py_phi}}
\end{figure}

\begin{figure}
\includegraphics[scale=0.5]{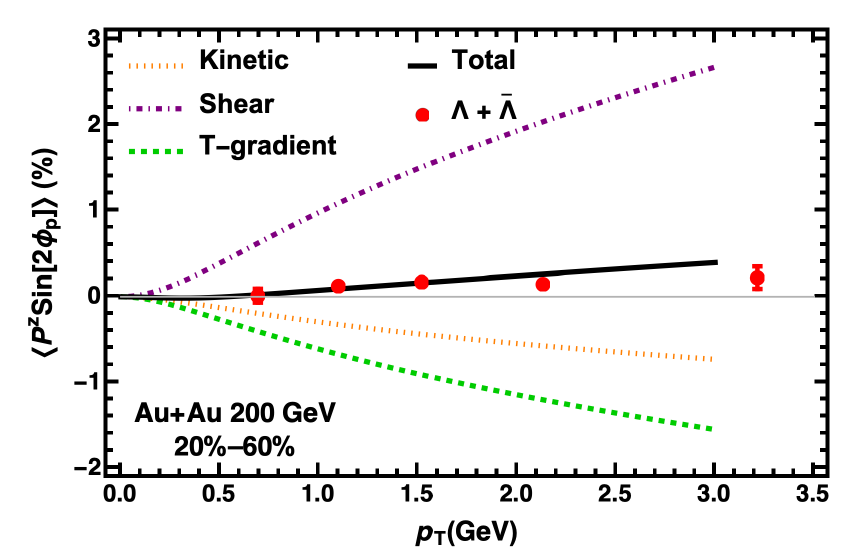}\includegraphics[scale=0.5]{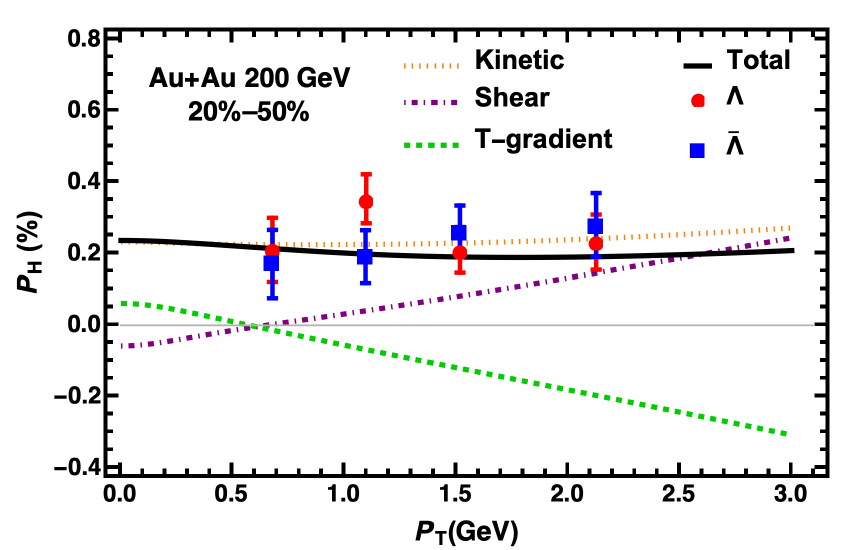}
\caption{The transverse momentum dependence of the second Fourier coefficient
of the longitudinal polarization, $\left<P^{z}\sin (2\phi_{p})\right>$
(left panel), and the local transverse polarization $P_{H}$ (right
panel), obtained from hydrodynamic simulation with the kinetic vorticity,
shear vorticity, and temperature-gradient sources in Au+Au collisions
at 200 GeV. The experimental data are taken
from STAR collaboration. \label{fig:p2z_py_pt}}
\end{figure}

\begin{figure}
\includegraphics[scale=0.5]{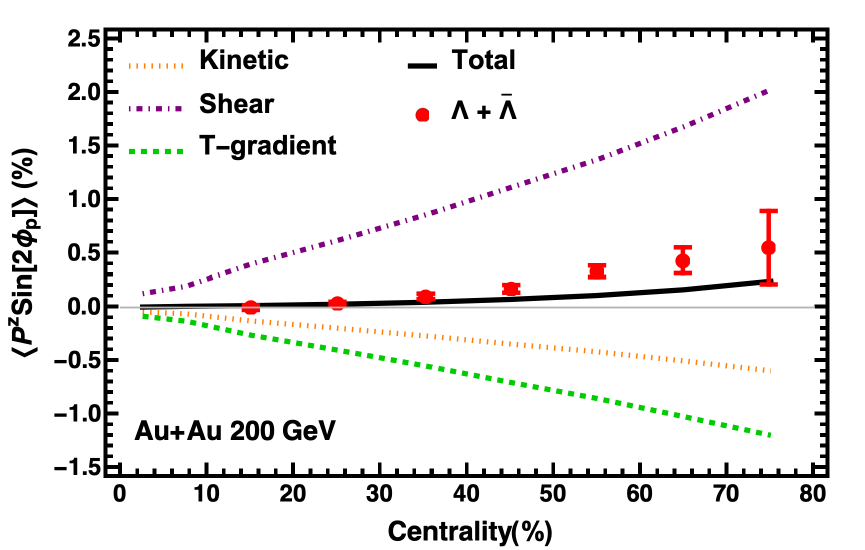}\includegraphics[scale=0.5]{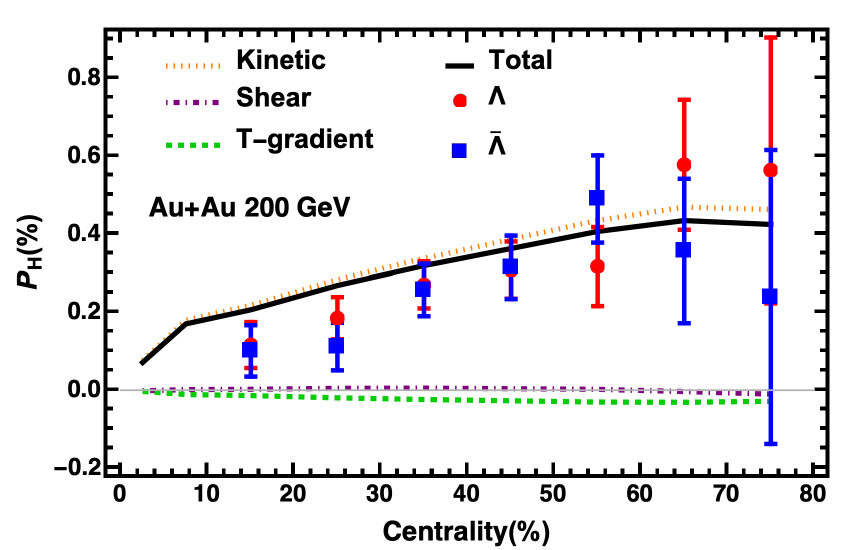}
\caption{The centrality dependence of the second Fourier coefficient of the
longitudinal polarization, $\left<P^{z}\sin (2\phi_{p}) \right>$
(left panel), and the out-of-plane polarization $P_{H}$ (right
panel), from hydrodynamic calculation that includes the kinetic vorticity,
shear vorticity, and temperature-gradient components in Au+Au collisions
at 200 GeV. The experimental data are obtained
from the STAR Collaboration. \label{fig:p2z_py_cent}}
\end{figure}

\begin{figure}
\includegraphics[scale=0.5]{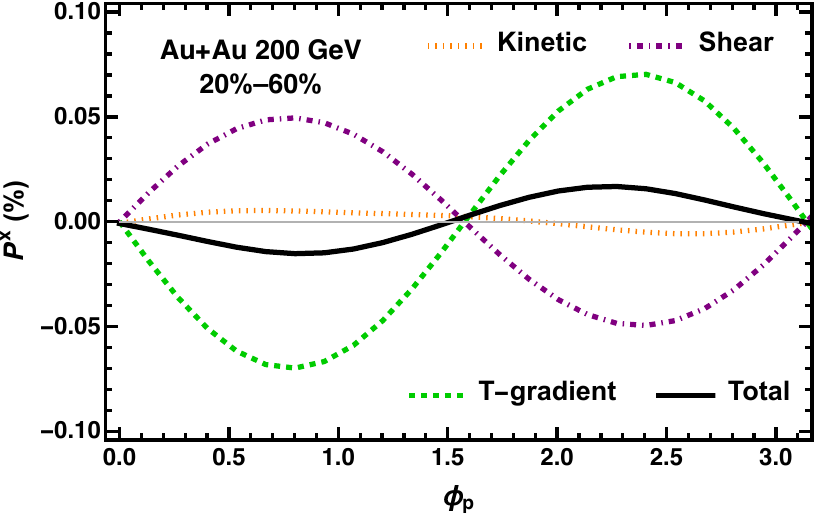}\hspace{0.5cm}\includegraphics[scale=0.5]{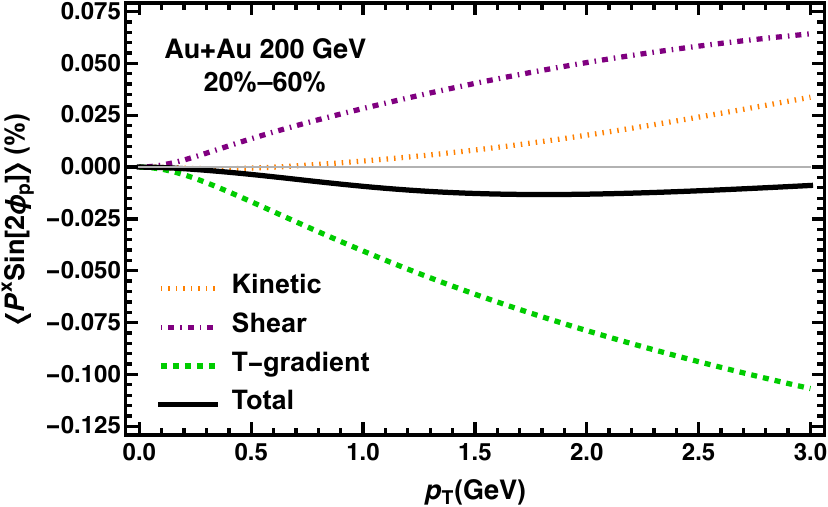}\\\includegraphics[scale=0.5]{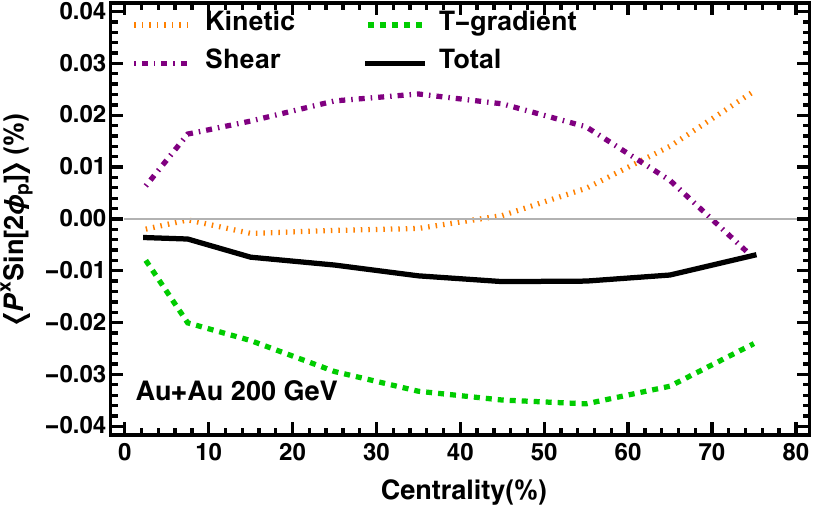}
\caption{First panel: hydrodynamic prediction for the in-plane polarization $P^{x}$ as
a function of the azimuthal angle; Second and third panels: hydrodynamic prediction
for the second Fourier coefficient, $\left<P^{x}\sin (2\phi_{p}) \right>$, as functions of $p_{T}$ and
centrality. The contributions from the kinetic vorticity, shear stress tensor
and temperature-gradient are shown separately. \label{fig:px_phi_pt_cent}}
\end{figure}

In Fig. \ref{fig:pz_py_phi}, we present the results for the azimuthal
angle ($\phi_{p}$) dependence of the longitudinal polarization
$P^{z}$ and the out-of-plane polarization $P_{H}$ from different
sources. It can be found that the hydrodynamic calculation can also
quantitatively describe the $P^{z}$ data, showing similar
results to those from previous hydro studies and from the blast wave model.
Among different sources, the shear-induced polarization gives the correct $\sin(2\phi _p)$
behavior compared to the data, and the shear term is the dominant contribution.
However, the polarizations by the kinetic vorticity and by the temperature-gradient
have the same sign but with different magnitudes. Interestingly, the magnitude
of the polarization from the temperature-gradient is even larger than
that from the kinetic vorticity. This indicates that the contribution
from the temperature gradient cannot be simply neglected, unlike what is assumed in
the blast wave model. Note that the magnitude of the data is much
smaller than those of the temperature-gradient, kinetic vorticity,
and shear contributions, it looks like the small signal is the result
of a sum over several large contributions and any small change
of one large contribution could flip the sign of the signal.

In Figs. \ref{fig:p2z_py_pt} and \ref{fig:p2z_py_cent}, we present
the $p_{T}$ and centrality dependence of the second
Fourier coefficient of the longitudinal polarization $\left<P^{z}\sin (2\phi_{p})\right>$,
and out-of-plane polarization $P_{H}$ in Au+Au collisions at 200 GeV.
The results demonstrate that the hydrodynamic simulation provides a quantitative description
of the experimental data.
For the longitudinal polarization $\left<P^{z}\sin (2\phi_{p})\right>$,
again, the shear contribution gives the correct slope in $p_T$ and centrality spectra,
while the temperature-gradient contribution gives an opposite slope, although
they have almost equal magnitude: two contributions are in competition.
For the out-of-plane polarization $P_{H}$, the hydrodynamic
calculation does not produce its $\phi _p$ dependence observed
in experiments as shown in Fig. \ref{fig:pz_py_phi}.
However, the magnitude of the total $P_{H}$ can be described
by the hydrodynamic model, as shown in Fig. \ref{fig:p2z_py_pt} and Fig. \ref{fig:p2z_py_cent}.
It can be found that $P_{H}$ mainly comes from the kinetic vorticity
contribution, which originates from the initial angular momentum.


The numerical results in Fig. \ref{fig:pz_py_phi}-\ref{fig:p2z_py_cent} show that
our hydrodynamic model can provide a description of the polarization data
and capture the gross features of the fireball's
dynamical evolution. Then we try to make a prediction for the in-plane
polarization $P^{x}$ as functions of $\phi _p$, $p_{T}$
and centrality in Fig. \ref{fig:px_phi_pt_cent}. Compared to
the prediction by the blast wave model, the hydrodynamic simulation result of $P^{x}$ shows
an overall opposite sign. This difference mainly arises from a significant contribution
from the temperature-gradient component in $P^{x}$, while the kinetic
and shear terms show the same sign as the blast wave prediction.
Similar to $P^{z}$, the magnitude of $P^{x}$ in total
is much smaller than those of the temperature-gradient and shear contributions,
it looks like the small signal is the result of the difference between
two large contributions.
The sign of $P^{x}$ can be flipped by any small change of one large contributions.
It has been shown in \citep{Lin:2024svh, Li:2025zbj} that, in an interacting system, the contribution of the velocity gradient and temperature gradient will be modified, which may enhance or diminish the result after cancellation.
The role played by the temperature-gradient in the in-plane polarization
will have to be elucidated through further experimental measurements.

\section{Summary}

We performed a comprehensive analysis of in-plane transverse polarization
in heavy-ion collisions, offering both analytical solutions and numerical
results. For the in-plane polarization, the hydrodynamic simulation
predicts an opposite sign compared to the blast wave model prediction,
mainly because of the contribution from the temperature-gradient component.
In the hydrodynamic simulation, similar to the longitudinal polarization,
the magnitude of the total result for the in-plane polarization is much smaller than those of
the temperature-gradient and shear contributions, i.e. the
small signal for the in-plane polarization is the result of the difference between two large contributions.
The sign of the signal can be easily flipped under any small change of either large contribution.
Future experiments may help to determine the impact of the temperature-gradient on the in-plane polarization.

\begin{acknowledgments}
A large part of this work was completed during the time when Q.W. was a visiting professor in
the nuclear theory group at McGill University.
This work is supported in part by the National Natural Science Foundation
of China (NSFC) under Grant No. 12135011, and in part by the Natural Sciences and Engineering Research Council of Canada (NSERC) [SAPIN-2020-00048, SAPIN-2024-00026].
\end{acknowledgments}

\appendix
\section{The coefficients in Eq. (\ref{eq:pol-xyz-phi-p}) \label{appendix-A}}
The explicit forms of $C^\omega_{z,n}$ with $n=0,1,\cdots,4$ in Eq. (\ref{eq:pol-xyz-phi-p}) are given as
\begin{align}
C_{z,0}^{\omega}= & -\frac{1}{2R^{2}}\rho_{0}\beta\left(\rho_{2}+\frac{1}{2}\epsilon\rho_{0}\right)rp_{T}^{2}\sinh\bar{\rho}\cosh\bar{\rho},\nonumber \\
C_{z,1}^{\omega}= & -\frac{1}{R^{2}}\rho_{0}\left(\rho_{2}+\frac{1}{2}\epsilon\rho_{0}\right)rp_{T}\cosh\bar{\rho}-\frac{1}{R}\left(3\rho_{2}+\frac{5}{2}\epsilon\rho_{0}\right)p_{T}\sinh\bar{\rho}\nonumber \\
 & +\frac{1}{R^{2}}\beta\rho_{0}\left(\rho_{2}+\frac{1}{2}\epsilon\rho_{0}\right)rp_{T}m_{T}\sinh^{2}\bar{\rho},\nonumber \\
C_{z,2}^{\omega}= & 4\rho_{2}\frac{1}{R}m_{T}\cosh\bar{\rho}-2\epsilon\frac{1}{r}m_{T}\sinh\bar{\rho},\nonumber \\
C_{z,3}^{\omega}= & \frac{1}{R^{2}}\rho_{0}\left(\rho_{2}+\frac{1}{2}\epsilon\rho_{0}\right)rp_{T}\cosh\bar{\rho}+\frac{1}{R}\left(-\rho_{2}+\frac{5}{2}\epsilon\rho_{0}\right)p_{T}\sinh\bar{\rho}\nonumber \\
 & -\frac{1}{R^{2}}\beta\rho_{0}\left(\rho_{2}+\frac{1}{2}\epsilon\rho_{0}\right)rp_{T}m_{T}\sinh^{2}\bar{\rho},\nonumber \\
C_{z,4}^{\omega}= & \frac{1}{2R^{2}}\rho_{0}\beta\left(\rho_{2}+\frac{1}{2}\epsilon\rho_{0}\right)rp_{T}^{2}\sinh\bar{\rho}\cosh\bar{\rho}.
\end{align}
The explicit forms of $C^{\xi}_{z,n}$ with $n=0,1,\cdots ,5$ are given as
\begin{align}
C_{z,0}^{\xi}= & \frac{1}{R}2\left(\rho_{2}+\epsilon\rho_{0}\right)p_{T}\cosh\bar{\rho}+\frac{1}{R^{2}}\rho_{0}\left(\rho_{2}+\frac{1}{2}\epsilon\rho_{0}\right)rp_{T}\sinh\bar{\rho}\nonumber \\
 & +\epsilon\frac{1}{r}p_{T}\sinh\bar{\rho}-\frac{3r}{2R^{2}}\beta\rho_{0}\left(\rho_{2}+\frac{1}{2}\epsilon\rho_{0}\right)p_{T}m_{T}\cosh\bar{\rho}\sinh\bar{\rho}\nonumber \\
 & +\frac{1}{R}\beta\left(\rho_{2}+\frac{1}{2}\epsilon\rho_{0}\right)m_{T}p_{T}\sinh^{2}\bar{\rho},\nonumber \\
C_{z,1}^{\xi}= & -\frac{1}{R^{2}}\rho_{0}\left(\rho_{2}+\frac{1}{2}\epsilon\rho_{0}\right)rm_{T}\cosh\bar{\rho}-\frac{1}{R}\left(3\rho_{2}+\frac{5}{2}\epsilon\rho_{0}\right)m_{T}\sinh\bar{\rho}\nonumber \\
 & +\frac{1}{R^{2}}\beta\rho_{0}\left(\rho_{2}+\frac{1}{2}\epsilon\rho_{0}\right)rp_{T}^{2}\cosh^{2}\bar{\rho}\nonumber \\
 & +\frac{1}{R^{2}}\beta\rho_{0}\left(\rho_{2}+\frac{1}{2}\epsilon\rho_{0}\right)rm_{T}^{2}\sinh^{2}\bar{\rho}\nonumber \\
 & -\frac{1}{R}\beta\left(\rho_{2}+\frac{1}{2}\epsilon\rho_{0}\right)p_{T}^{2}\cosh\bar{\rho}\sinh\bar{\rho},\nonumber \\
C_{z,2}^{\xi}=& 0\nonumber,\\
C_{z,3}^{\xi}= & \frac{1}{R^{2}}\rho_{0}\left(\rho_{2}+\frac{1}{2}\epsilon\rho_{0}\right)rm_{T}\cosh\bar{\rho}+\frac{1}{R}\left(-\rho_{2}+\frac{5}{2}\epsilon\rho_{0}\right)m_{T}\sinh\bar{\rho}\nonumber \\
 & -\frac{1}{2R^{2}}\beta\rho_{0}\left(\rho_{2}+\frac{1}{2}\epsilon\rho_{0}\right)rp_{T}^{2}\cosh^{2}\bar{\rho}\nonumber \\
 & -\frac{1}{R^{2}}\beta\rho_{0}\left(\rho_{2}+\frac{1}{2}\epsilon\rho_{0}\right)rm_{T}^{2}\sinh^{2}\bar{\rho}\nonumber \\
 & +\frac{1}{2R}\left(\rho_{2}+\frac{1}{2}\epsilon\rho_{0}\right)\beta p_{T}^{2}\cosh\bar{\rho}\sinh\bar{\rho},\nonumber \\
C_{z,4}^{\xi}= & \frac{1}{R}2\left(\rho_{2}-\epsilon\rho_{0}\right)p_{T}\cosh\bar{\rho}-\frac{1}{R^{2}}\rho_{0}\left(\rho_{2}+\frac{1}{2}\epsilon\rho_{0}\right)rp_{T}\sinh\bar{\rho}\nonumber \\
 & +\epsilon\frac{1}{r}p_{T}\sinh\bar{\rho}-\frac{1}{R}\beta\left(\rho_{2}+\frac{1}{2}\epsilon\rho_{0}\right)m_{T}p_{T}\sinh^{2}\bar{\rho}\nonumber \\
 & +\frac{3}{2R^{2}}\beta\rho_{0}\left(\rho_{2}+\frac{1}{2}\epsilon\rho_{0}\right)rp_{T}m_{T}\cosh\bar{\rho}\sinh\bar{\rho},\nonumber \\
C_{z,5}^{\xi}= & -\frac{1}{2R^{2}}\beta\rho_{0}\left(\rho_{2}+\frac{1}{2}\epsilon\rho_{0}\right)rp_{T}^{2}\cosh^{2}\bar{\rho}\nonumber \\
 & +\frac{1}{2R}\beta\left(\rho_{2}+\frac{1}{2}\epsilon\rho_{0}\right)p_{T}^{2}\cosh\bar{\rho}\sinh\bar{\rho}.
\end{align}

\bibliographystyle{unsrt}
\addcontentsline{toc}{section}{\refname}\bibliography{referances-polx}

\end{document}